\documentclass[preprint]{aastex}

\usepackage[onecolumn]{emulateapj5}

\newcommand{\himpc}{{\hbox {$\,h^{-1}$}{\rm Mpc}} }
\newcommand{\himpcz}{{\hbox {$\,h^{-1}{\rm Mpc}_z$}} }
\newcommand{\bfa}{{\mbox{\boldmath $a$}}}
\newcommand{\bfx}{{\mbox{\boldmath $x$}}}

\newcommand{\bfs}{{\mbox{\boldmath $s$}}}

\newcommand{\bfC}{{\mbox{\boldmath $C$}}}
\newcommand{\bfF}{{\mbox{\boldmath $F$}}}
\newcommand{\bfth}{{\mbox{\boldmath $\theta$}}}
\newcommand{\sbfk}{{\mbox{\scriptsize\boldmath $k$}}}
\newcommand{\sbfx}{{\mbox{\scriptsize\boldmath $x$}}}

\newcommand{\omm}{\Omega_{\rm M}}
\newcommand{\oml}{\Omega_{\rm \Lambda}}
\newcommand{\fbaryon}{\Omega_{\rm B}/\Omega_{\rm M}}

\begin{document}

\renewcommand{\theequation}{\mbox{\rm
{\arabic{section}.\arabic{equation}}}} 


\title{Cosmological Parameters from Redshift-Space Correlations}

\author{Takahiko Matsubara}
\affil{Department of Physics and Astrophysics, 
	Nagoya University,
	Chikusa, Nagoya 464-8602, Japan}

\and

\author{Alexander S. Szalay}
\affil{Department of Physics and Astronomy, 
        The Johns Hopkins University,
        Baltimore, MD 21218}

\email{taka@a.phys.nagoya-u.ac.jp, szalay@jhu.edu}

\begin{abstract}

We estimate how clustering in large-scale redshift surveys can
constrain various cosmological parameters. Depth and sky coverage of
modern redshift surveys are greater than ever, opening new
possibilities for statistical analysis.  We have constructed a novel
maximum likelihood technique applicable to deep redshift surveys of
wide sky coverage by taking into account the effects of both curvature
and linear velocity distortions.  The Fisher information matrix is
evaluated numerically to show the bounds derived from a given redshift
sample.  We find that intermediate-redshift galaxies, such as the
Luminous Red Galaxies (LRGs) in the Sloan Digital Sky Survey, can
constrain cosmological parameters, including the cosmological
constant, unexpectedly well. The importance of the dense as well as
deep sampling in designing redshift surveys is emphasized.

\end{abstract}



\setcounter{equation}{0}
\section{Introduction}
\label{sec1}

Recently, independent cosmological observations have turned out to be
consistent with a canonical model of the structure formation based on
the assumption that the main ingredients of our universe are the cold
dark matter (CDM), and the dark energy component such as a
cosmological constant. Introducing the unknown dark matter and the
peculiar dark energy could have been thought of as somewhat artificial
at the time when observations were not constraining enough, now this
model has become more and more convincing with remarkable
observational support.

To make a cosmological model truly convincing, it is important to make
sure that any possible observations are consistent with that model.
This can be accomplished by determining cosmological parameters as
precisely as possible from each observation, and then comparing the
results with one another. So far the individual observations were only
able to constrain the parameters into some region in a multi-dimensional
parameter space. Therefore, it is quite common to combine different
observations to obtain the tight constraint on the parameters. A
striking example is a combination of fluctuations in cosmic microwave
background \citep{deb00} and a Hubble diagram of supernova Ia
\citep{rie98,per99} which is considered to be the prominent evidence of
the current canonical model.

However, one can test the consistency of the theory only when each
observation can independently constrain the parameters in a
sufficiently small region. In this paper, we explore the capabilities
of redshift surveys in this respect. In redshift surveys, there are
several degrees of freedom in planning survey strategies: the area and
the depth of a survey, and the choice of specific targets for the
spectroscopic measurements. Each of these affects the analysis in a
different and complex fashion. Thus, it is important to estimate the
impact of a given strategy on the cosmological analysis of the data.

In this paper, we apply our considerations to the Sloan Digital Sky
Survey \citep[SDSS,][]{yor00}. It is not only the largest ongoing
redshift survey, but it also contains three different types of
spectroscopic targets within the same area of the sky that can be used
to infer cosmological information.

In deep, wide redshift surveys like SDSS, the apparent clustering
properties in observable redshift space are complicated because of the
redshift-evolution and redshift-distortion effects. In linear regime,
with a standard initial Gaussian density field, the correlation
function $\xi(r)$ or the power spectrum $P(k)$ completely
characterizes the statistical property of the density field. Spatial
homogeneity and isotropy in real space guarantees these to be
one-dimensional functions of the scale, $r$ or $k$. So it is usual to
estimate the real-space power spectrum with certain correction methods
or ansatz from observed redshift-space data, in order to compare with
the theoretical prediction of various cosmological models
\citep[e.g.,][]{dav83,fel94}. 

However, the redshift-evolution and redshift-distortion also have
invaluable information on cosmology. The redshift-evolution of the
density fluctuations depends on the growth factor which is a function
of the density parameter $\omm$ and the normalized cosmological
constant parameter $\oml$. It is also a function of the galaxy bias
$b$ which contains the information on the unknown galaxy formation
process. The redshift-distortion depends on the density-velocity
relation which is a function of $\omm^{0.6}/b$ \citep{kai87}, and also
on $\omm$ and $\oml$ through geometrical distortions
\citep{alc79,bal96,mat96}. 

When our main purpose is not only determining the power spectrum of
initial fluctuations, but also constraining above parameters, it is
desirable to directly analyze the data to obtain cosmological
information without any intermediate statistical quantities like the
power spectrum. Since the apparent clustering in redshift space is
quite different from that in real space, we can take advantage of the
maximum-likelihood method to directly constrain cosmological models
from the apparent clustering data in redshift space.

In our previous work \citep{mat01}, we focused on how near-future
redshift surveys can constrain the cosmological constant by the
maximum-likelihood method. Since the redshift surveys become larger
and larger, even the multi-dimensional parameter space can be
constrained by redshift surveys alone. Combining with other
cosmological observations like the cosmic microwave background
fluctuations, Hubble diagram of the type Ia supernova, and so on, one
can constrain the parameters more accurately. However, independent
observational constraints can serve as a consistency check of the
fundamental cosmological model. In this respect, we investigate how
one can constrain the multi-dimensional parameter space of the
standard CDM model with cosmological constant from redshift surveys,
extending our previous work.

\section{Maximum-likelihood Method for the Apparent Redshift-space
Clustering}
\label{sec2}

In maximum-likelihood methods, we directly deal with the density
fluctuations in redshift space. In order to do this, first we place
smoothing cells in redshift space and consider the number of objects
$n_i$ in each cell $i$. Each count represents the smoothed density
field contaminated by shot noise. For given mean number densities,
$\bar{n}_i = \langle n_i \rangle$, the density fluctuations $a_i =
n_i/\bar{n}_i - 1$ are the fundamental quantities to be analyzed. In
the following, the most important statistical quantity is the
correlation matrix, $C_{ij} = \langle a_i a_j \rangle$. This matrix is
related to a continuous two-point correlation function in redshift
space $\xi(\bfs, \bfs') = \langle \delta(\bfs) \delta(\bfs') \rangle$,
where $\delta(\bfs)$ is the underlying continuous density contrast
which survey objects trace. The relation between the correlation
matrix and the correlation function is given by
\begin{eqnarray}
   C_{ij} =
   \int d^3 s\, d^3 s'\, W^{\rm (s)}(\bfs_i - \bfs)
   W^{\rm (s)}(\bfs_j - \bfs')\,
   \xi(\bfs,\bfs') +
   \int d^3 s\,
   \frac{W^{\rm (s)}(\bfs_i - \bfs) W^{\rm (s)}(\bfs_j - \bfs)}
      {\bar{n}(\bfs)},
\label{eq1-1}
\end{eqnarray}
where $W^{\rm (s)}(\bfs)$ is the smoothing kernel in redshift space,
normalized as $\int d^3 s W^{\rm (s)}(\bfs) = 1$ and $\bar{n}(\bfs)$
is the mean number density. The last term represents contribution from
the shot noise effect. In linear regime, the two-point correlation
function in redshift space is analytically given, including high
redshift effect.

Although the general form is fairly complicated \citep{mat00}, the
distant-observer approximation \citep{kai87, mat96}, which is
applicable in many cases, simplify the analysis below. The
corresponding formula \citep{mat96,mat00} is given by
\begin{eqnarray}
   \xi(\bfs_1,\bfs_2) =
   b(z_1) b(z_2) D(z_1) D(z_2)
   \int\frac{d^3k}{(2\pi)^3}
   e^{i\sbfk\cdot(\sbfx_1 - \sbfx_2)}
   \left[1 + \beta(z_1) \frac{{k_\parallel}^2}{k^2} \right]
   \left[1 + \beta(z_2) \frac{{k_\parallel}^2}{k^2} \right] P(k),
\label{eq1-2}
\end{eqnarray}
where $z_1$ and $z_2$ are the redshifts of the two points,
$k_\parallel$ is the line-of-sight component of the wave number,
$P(k)$ is the linear mass power spectrum at $z=0$, $D(z)$ is the
linear growth rate normalized as $D(0)=1$, $b(z)$ is the bias factor
at redshift $z$, and $\beta(z)$ is the redshift distortion parameter,
which is accurately fitted by the redshift-dependent mass density
parameter $\omm(z)$ and normalized cosmological constant $\oml$ as
$\beta(z) \simeq \{\omm^{4/7}(z) + \oml(z)[1 + \omm(z)/2]/70\}/b(z)$
\citep{lah91,lig90}. The vectors $\bfx_1$ and $\bfx_2$ are the
comoving positions of the two points which are labeled by $\bfs_1$ and
$\bfs_2$ in redshift space \citep[see][]{mat96}.

Except for the nearby universe, the comoving distance is not
proportional to the redshift so that the smoothing kernel $W^{\rm
(s)}$ in redshift space is distorted along the line of sight. In a
distant-observer regime, a line-of-sight component $s_\parallel$ and a
transverse component $s_\perp$ of redshift-space separation $\bfs$ are
given by
\begin{eqnarray}
   s_\parallel = H(z) x_\parallel,
   \qquad
   s_\perp = \frac{z}{d_A(z)} x_\perp,
\label{eq1-2-1}
\end{eqnarray}
where $x_\parallel$ and $x_\perp$ are corresponding line-of-sight and
transverse components of real-space separation, respectively, and $z$
is the mean redshift. We have chosen the unit system $H_0 = 1$ here
for simplicity. This unit system simplifies the equations in this
section. We mention that another unit system for length is introduced
in \S\ref{sec4} for a practical reason which is explained there. In
the above transform, the redshift-dependent Hubble's parameter and the
comoving angular diameter distance are explicitly given by
\begin{eqnarray}
&&
   H(z) = \sqrt{\omm (1+z)^3 + (1-\omm-\oml) (1+z)^2 + \oml},
\label{eq1-2-2a}\\
&&
   d_A(z) =
   \left\{
   \begin{array}{ll}
      \displaystyle
      (-K)^{-1/2} \sinh\left[ (-K)^{1/2} \chi(z) \right] & (K<0),\\
      \displaystyle
      \chi(z) & (K=0),\\
      \displaystyle
      K^{-1/2} \sin\left[ K^{1/2} \chi(z) \right] &(K>0),
   \end{array}
   \right.
\label{eq1-2-2b}
\end{eqnarray}
where 
\begin{eqnarray}
   K = 1 - \omm - \oml,
   \qquad
   \chi(z) = \int_0^z \frac{dz'}{H(z')},
\label{eq1-2-3}
\end{eqnarray}
are the curvature of the universe and comoving distance to the
redshift $z$. Thus, the transform from the comoving space to the
redshift space is the function of $\omm$ and $\oml$

As a result of the anisotropic transform of equation (\ref{eq1-2-1}),
the spherical kernel in redshift space, for example, ends up with
oscillating two-dimensional numerical integration \citep{mat01} in
evaluating the correlation matrix $C_{ij}$. Unfortunately, such
numerical integration increases CPU time and decreases accuracy. This
is crucial for the analysis of the real data, because the position of
kernels can hardly be regular so that one has to calculate the
correlation matrix pair by pair. The shape of the kernel is not
important for largely separated pairs, but it affects nearby and
identical pairs.

Although one can not exactly know what is the distortion unless the
parameters $\omm$ and $\oml$ are fixed, it turns out to be a good
approximation to set an approximate spherical kernel in real space
with approximate values of these parameters. In fact, we found that
the latter approximation is better than doing the exact numerical
integration which inevitably has integration errors. When the
parameters $\omm$ and $\oml$ should also be determined in real
analyses, one can iterate the analysis to use better estimates for
those parameters, but approximate values are good enough. Thus,
suppose that we can approximately set a spherical kernel in real
space, so that we have $W^{\rm (s)}(\bfs) =J^{-1} W_R(|\bfx|)$, where
$J = \partial(\bfs)/\partial(\bfx)$ is the Jacobian from $x$-space to
$s$-space, and $R$ is the (effective) smoothing radius. In the
distant-observer approximation, $J(z) = z^2 H(z)/{d_A}^2(z)$.

This approximation dramatically simplifies the evaluation of the
correlation matrix. In fact, when the kernels are spherical in real
space, at least approximately, equations (\ref{eq1-1}) and
(\ref{eq1-2}) give
\begin{eqnarray}
&&
   C_{ij} = 
   b_i b_j D_i D_j \int \frac{d^3k}{(2\pi)^3}
   e^{i\sbfk\cdot(\sbfx_i - \sbfx_j)}
   \left[1 + \beta_i \frac{{k_\parallel}^2}{k^2} \right]
   \left[1 + \beta_j \frac{{k_\parallel}^2}{k^2} \right]
   W^2(kR) P(k)
\nonumber\\
&& \qquad\qquad\qquad +\,
   \frac{1}{\sqrt{{J_i J_j \bar{n}_i \bar{n}_j}}}
   \int \frac{d^3k}{(2\pi)^3} e^{i\sbfk\cdot(\sbfx_i - \sbfx_j)}
   W^2(kR),
\label{eq1-3}
\end{eqnarray}
where $b_i = b(z_i)$, $D_i = D(z_i)$, $J(z_i) = J_i$, etc., and
$W(kR)$ is the three-dimensional Fourier transform of the spherical
kernel $W_R(|\bfx|)$. In our notation, $\bar{n}_i$ is the mean number
density in redshift space so that $J_i \bar{n}_i$ is the mean number
density in real space. We assume that the variation of the selection
function within a kernel is negligible. The above equation means that
the distant-observer redshift-distortion is commutable with spherical
smoothing in real space, comparing with equation (\ref{eq1-2}) besides
shot-noise term. Angular integration of the first term can be
analytically performed \citep[see][]{ham92,mat96,mat00}, resulting in
\begin{eqnarray}
&& C_{ij} =
   b_i b_j D_i D_j
   \left\{
      \left[
         1 + \frac13 (\beta_i + \beta_j) + \frac15 \beta_i \beta_j
      \right]
      P_0(\mu_{ij}) \xi_0(x_{ij},R)
   \right.
\nonumber\\
&& \quad\qquad\qquad\qquad -\,
   \left.
      \left[
         \frac23 (\beta_i + \beta_j) + \frac47 \beta_i \beta_j
      \right]
      P_2(\mu_{ij}) \xi_2(x_{ij},R)
    +
      \frac{8}{35} \beta_i \beta_j
      P_4(\mu_{ij}) \xi_4(x_{ij},R)
   \right\}
\nonumber\\
&& \qquad\qquad\qquad +\,
   \frac{1}{\sqrt{{J_i J_j \bar{n}_i \bar{n}_j}}}
   \int \frac{k^2dk}{2\pi^2} j_0(kx_{ij}) W^2(kR),
\label{eq1-4}
\end{eqnarray}
where $\beta_i = \beta(z_i), \beta_j = \beta(z_j)$, $x_{ij} = |\bfx_i
- \bfx_j|$ is the comoving separation between the centers of the
kernels, $\mu_{ij} = x_{ij\parallel}/x_{ij}$ is the direction-cosine
of the separation $x_{ij}$ relative to the line of sight, $j_l$ is the
spherical Bessel function, and $P_l(\mu)$ is the Legendre polynomial.
We also define the quantity,
\begin{eqnarray}
   \xi_{l}(x,R) = \int_0^\infty \frac{k^2dk}{2\pi^2}
   j_{l}(kx) W^2(kR) P(k).
\label{eq1-5}
\end{eqnarray}

We need to numerically integrate the equation (\ref{eq1-5}). Once we
make a table of the integrated results for various $x$ and $R$ of
interest, we can use this table to interpolate $\xi_l$, which makes
the evaluation of the correlation matrix $C_{ij}$ a really fast
operation. This method is very effective because the number of
elements of a correlation matrix is the square of the number of cell
sites $(N_{\rm cell})^2$, where $N_{\rm cell} \sim 1,000$--$10,000$,
or more. We tried exact 2-dimensional integrations for identical and
nearby cell-pairs, and found that they agree with the above
approximation within the integration errors, even though the assumed
$\omm$ and $\oml$ are different from the true values by a factor of
about 0.5.

The shot noise term is diagonal when the kernels do not overlap each
other, as one can see from the equation (\ref{eq1-1}). Using a
finite-volume kernel without any overlapping region, the last term of
equation (\ref{eq1-1}) or (\ref{eq1-4}) is given by
\begin{eqnarray}
   \frac{\delta_{ij}}{\bar{n}_i}
   \int d^3 s \left[W^{\rm (s)}(\bfs)\right]^2
   =
   \frac{\delta_{ij}}{J_i \bar{n}_i}
   \int \frac{k^2dk}{2\pi^2} W^2(kR).
\label{eq1-5-1}
\end{eqnarray}
The top-hat kernel, $W_R(x) = 3\Theta(R-x)/(4\pi R^3)$, where $\Theta$
is the step function, is the most popular example of the finite-volume
kernel, in which case the above shot noise term simply reduces to
$\delta_{ij}/\bar{N}_i$, where $\bar{N}_i = 4\pi R^3 J_i \bar{n}_i/3$ is
the expected mean number count in a cell.

Although we only use the top-hat kernel in this paper, it is also
useful to consider another possibility as digression, since this fast
numerical approximations will also be used in the analysis of the real
data. The top-hat kernel is not a smooth function, thus it is
sometimes desirable to have an smoother kernel. An example of the
smoother, yet finite-volume kernels is the Epanechnikov kernel, which
is defined by $W_R(x) = 15(1-x^2/R^2)\Theta(R-x)/(8\pi R^3)$. While
the top-hat kernel is discontinuous on the edge $x=R$, the
Epanechnikov kernel is continuous. The derivative is still
discontinuous on the edge. One can also use the $m$-weight
Epanechnikov kernel defined by
\begin{eqnarray}
   K_m(x;R) = 
   \frac{(2m+3)!!}{ m!\ 2^{m+2} \pi R^3}
   \left(1 - \frac{x^2}{R^2}\right)^m \Theta(R-x).
\label{eq1-6}
\end{eqnarray}
The top-hat kernel and the Epanechnikov kernel are the special cases
of this general Epanechnikov kernel with $m=0$ and $m=1$,
respectively. The $(m-1)$-derivative of the $m$-weight Epanechnikov
kernel is continuous on the edge. The larger the rank in the general
Epanechnikov kernel is, the more regular it is on the edge. However,
the effective volume of the kernel is smaller for a larger-weight
kernel when a smoothing length $R$ is fixed. Taking limits
$m\rightarrow\infty, R\rightarrow\infty$ with $m^{-1/2}R$ fixed, one
can show that the higher-weight Epanechnikov kernel reduces to a
Gaussian kernel with a variance $R/\sqrt{2m}$. The Fourier window
function for the $m$-weight Epanechnikov kernel has a simple form,
\begin{eqnarray}
   W_m(kR) \equiv
   \int_0^\infty d^3 x e^{-i\sbfk\cdot\sbfx} K_m(x;R) =
   \frac{(2m+3)!!\ j_{m+1}(kR)}{(kR)^{m+1}}.
\label{eq1-7}
\end{eqnarray}
The larger the rank $m$ is, the faster the window function drops off
with the wavenumber $k$, which reflects the smoothness on the edge of
the kernel. It is desirable to have a degree of freedom to choose an
appropriate rank $m$, because one can optimize the analysis
depending on the nature of an individual data set.

The observational data is the set of discrete density fluctuations
$a_i$. The cosmological models are characterized by a given set of
parameters $(\theta_\alpha)$. Besides $\omm$, $\oml$, and $b$, which
is explicitly appeared in the expression of the correlation matrix,
the power spectrum contains several other parameters when we
characterize it by a standard cold dark matter model. The latter
parameters are the normalization $\sigma_8$, baryon mass fraction
$f_{\rm B}\equiv\fbaryon$, spectral index $n$, and Hubble constant
$H_0$, etc.

According to the Bayes' rule, the direct likelihood function of the
cosmological parameters for a data set is given by
\begin{eqnarray}
   {\cal L}(\bfth | \bfa) =
   \frac{{\cal L}(\bfa | \bfth){\cal L}(\bfth)}{{\cal L}(\bfa)},
\label{eq1-8}
\end{eqnarray}
where ${\cal L}(\bfth | \bfa)$ indicates the conditional likelihood
function of the parameter set $\bfth = (\theta_1, \theta_2, \ldots,
\theta_M)$ given the date set $\bfa = (a_1, a_2, \ldots, a_N)$, ${\cal
L}(\bfth)$ indicates the prior likelihood function of the parameter
set $\bfth$, and so on. Normalization determines the denominator of
the right hand side, and the prior likelihood ${\cal L}(\bfth)$ comes
from the other observations, or just a constant if we do not combine
our prier knowledge on the cosmological parameters. The remaining
factor ${\cal L}(\bfa | \bfth)$ is the distribution function of the
data set given a cosmological model. In linear regime where the
density fluctuations are considered to be random Gaussian, this
function is simply the multi-variate Gaussian distribution function:
\begin{eqnarray}
   {\cal L}(\bfa | \bfth) =
   \frac{1}{\sqrt{(2\pi)^N \det\bfC(\bfth)}}
   \exp\left(-\frac12 \bfa^{\rm T} \bfC^{-1} \bfa \right),
\label{eq1-9}
\end{eqnarray}
where $\bfC^{-1}$ is the inverse matrix of the correlation matrix
$\bfC = (C_{ij})$. The matrix $\bfC$ is the function of cosmological
parameters $\bfth$ as noted above. Thus, calculation of the likelihood
function is reduced to the calculation of equation (\ref{eq1-4}) for a
given survey geometry and selection function for various cosmological
models.

\setcounter{equation}{0}
\section{Evaluation of the Fisher Information Matrix}
\label{sec3}

The Fisher information matrix, 
\begin{eqnarray}
   F_{\alpha\beta}(\bfth) =
   \left\langle
     \frac{\partial\ln{\cal L}(\bfa | \bfth)}{\partial\theta_\alpha}
     \frac{\partial\ln{\cal L}(\bfa | \bfth)}{\partial\theta_\beta}
   \right\rangle
   = -
   \left\langle \frac{\partial^2 \ln {\cal L}(\bfa | \bfth)}
         {\partial\theta_\alpha\partial\theta_\beta}
   \right\rangle,
\label{eq1-10}
\end{eqnarray}
where $\langle\cdots\rangle = \int {\cal L}(\bfa | \bfth) \cdots d^N
a$ represents the averaging over the possible data realization given
the model parameters, is the key quantity in estimation theory,
because the Cram\'er-Rao theorem \citep{ken69,the92} states that an
unbiased estimator constrains the model parameters with a minimum
variance
\begin{eqnarray}
   \left\langle \Delta\theta_\alpha \Delta\theta_\beta \right\rangle
   \mbox{``} \geq \mbox{''}
   \left( F^{-1} \right)_{\alpha\beta},
\label{eq1-11}
\end{eqnarray}
where $F^{-1}$ is the inverse matrix of the Fisher matrix and
$\Delta\theta_\alpha = \hat{\theta}_\alpha - \theta_\alpha$ is the
deviation of the estimated value $\hat{\theta}_\alpha$ from the
``true'' value $\theta_\alpha$. In the above equation, the notation
$\mbox{``} \geq \mbox{''}$ means that the matrix whose
$\alpha\beta$-elements are given by $\left\langle \Delta\theta_\alpha
\Delta\theta_\beta \right\rangle - \left( F^{-1}
\right)_{\alpha\beta}$ is positive semidefinite. Moreover, the
equation (\ref{eq1-11}) is satisfied with equality if one uses the
maximum likelihood estimate, which is obtained by a maximization of
the likelihood function ${\cal L}(\bfa | \bfth)$ as in the previous
section:
\begin{eqnarray}
   \left\langle \Delta\theta_\alpha \Delta\theta_\beta \right\rangle
   =
   \left( F^{-1} \right)_{\alpha\beta}
   \qquad \mbox{(maximum likelihood estimate)}.
\label{eq1-12}
\end{eqnarray}
Thus, the inverse of the Fisher matrix gives us the expected
constraint on the model parameters $\bfth$. A fiducial model for the
``true'' values of $\bfth$ are required to calculate the Fisher
matrix.

In the linear regime where the distribution function is given by
equation (\ref{eq1-9}), the Fisher matrix is analytically reduced to
the following form \citep[see, e.g.,][]{vog96,teg97}:
\begin{eqnarray}
   F_{\alpha\beta} =
   \frac12 {\rm Tr}
   \left(
      \bfC^{-1} \frac{\partial\bfC}{\partial\theta_\alpha}
      \bfC^{-1} \frac{\partial\bfC}{\partial\theta_\beta}
   \right)
\label{eq1-13}
\end{eqnarray}
Using this equation, the calculation of the Fisher matrix is
straightforward once the calculation of the correlation matrix for a
given model is established in the previous section.

The ellipsoid defined by an equation $\Delta\bfth^{\rm T} \bfF
\Delta\bfth = \nu^2$ defines the {\em concentration ellipsoid} which
is interpreted as error bounds for a given observation. If the
parameter likelihood function ${\cal L}(\bfth | \bfa)$ is Gaussian,
the concentration ellipsoid with a fixed value of $\nu$ represents the
$\nu\sigma$ confidence level when one uses the maximum likelihood
estimate. Although in general the parameter likelihood function is not
Gaussian, the concentration ellipsoids still give the rough idea of
expected error bounds with a given confidence level of $\nu$.

When the parameter space is just two-dimensional, one can easily plot
the concentration ellipses. The presentation of higher-dimensional
space is not easy. We have at most seven parameters in our model:
$(\omm, \oml, b, f_{\rm B}, h, \sigma_8, n)$. To depict the higher
dimensional concentration ellipsoids, we can plot all the possible
two-dimensional parameter spaces which are marginalized over all the
other parameters. This gives the idea how one can impose the
constraints in the multi-dimensional parameter space. Although the
form of the parameter likelihood function is not known, we can assume
it is approximated by a multi-variate Gaussian function so that ${\cal
L}(\bfth | \bfa) \sim \exp(-\frac12 \Delta\bfth^{\rm T} \bfF
\Delta\bfth)$. In which case, the marginalization over the other
parameters but two can be performed. For example, from the standard
Gaussian integration, marginalized 2-dimensional likelihood function
for two parameters $\theta_\alpha$ and $\theta_\beta$ is given by
\begin{eqnarray}
  {\cal L}(\theta_\alpha, \theta_\beta | \bfa)
  \sim
  \exp\left\{
    -\frac12
    \left(\Delta\theta_\alpha\ \Delta\theta_\beta\right)
    \left[
      \begin{array}{cc}
         \left(F^{-1}\right)_{\alpha\alpha} &
         \left(F^{-1}\right)_{\alpha\beta} \\
         \left(F^{-1}\right)_{\alpha\beta} &
         \left(F^{-1}\right)_{\beta\beta}
      \end{array}
    \right]^{-1}
    \left(
      \begin{array}{c}
        \Delta\theta_\alpha \\
        \Delta\theta_\beta
      \end{array}
    \right)
   \right\},
\label{eq1-14}
\end{eqnarray}
up to normalization factor, where $(F^{-1})_{\alpha\beta}$ indicates
the $\alpha\beta$-element of the inverse matrix of the original
higher-dimensional Fisher matrix, and so on. Therefore, plotting the
concentration ellipses in the marginalized parameter space is
straightforward: they are given by contours of the equation
(\ref{eq1-14}). Generally, concentration ellipsoids in the
marginalized parameter sub-space are deduced from the sub-matrix of
the full inverse Fisher matrix.

\section{Results for the SDSS geometry}
\label{sec4}

The largest redshift map which will be available in next several years
is the SDSS survey. Therefore, it is of great interest how the
maximum-likelihood method described in \S \ref{sec2} can constrain
various cosmological parameters. We have seen in \S \ref{sec3} that it
is straightforward to estimate the expected constraints on parameters
given the survey geometry. We examine the maximum-likelihood method with
the SDSS geometry as a typical application of our method.

Redshift-space observations in the SDSS survey consist of three
samples of objects \citep{yor00}. The first sample is the set of
approximately 1,000,000 ``Main Galaxies'', which are selected by
limiting an $r'$-band magnitude. The redshift of main galaxies extends
up to $z\sim 0.2$. The second sample is the set of approximately
120,000 ``Luminous Red Galaxies'' (LRG's), which are selected on the
basis of color and magnitude \citep{eis01}. Because of the clever
selection criteria, the LRG's are essentially a distance-limited
sample whose redshift extends up to $z\sim 0.45$. The third sample is
the set of approximately 100,000 quasars, which are selected from
color-color diagrams. The redshift of the quasar sample extends up to
$z\sim 3$.

In evaluating Fisher information matrices, the precise geometry of the
survey is not needed, and one can approximate the survey geometry to
reduce the numerically intense procedure of inverting huge matrices.
We setup a generic cubic box with 1,000 cells for each samples. The
redshift of the box is set to be a ``mean'' redshift of each sample:
$z=0.1$ for main galaxies, $z=0.3$ for LRG's, and $z=1.8$ for quasars.
The precise value of this mean redshift is not a decisive factor in
the Fisher matrix evaluations. The finiteness of the spatial volume
and the number density of the objects are the main source of the
statistical uncertainty, i.e., cosmic variance and shot noise. From
the sky coverage of the three samples of the SDSS spectroscopic
survey, which is about pi-steradian, and from the redshift
distribution of each samples, one can obtain expected spatial volumes
and number densities for each sample.

Although the unit system $H_0=1$ used in the previous sections
simplifies the equations, it is not convenient in mentioning the
clustering scales which become small figures in this system. Thus, we
use a different unit system when we have to mention the length scale
in the following paragraphs. Following \citet{mat01}, we adopt a
coordinate system in which radial distance $s$ to the objects is
measured by redshift times Hubble length: $s = cz/H_0 = (2997.9
\himpc) \times z$, directly observable quantity, instead of the
comoving distance. The latter depends on cosmological models through
parameters $\omm$ and $\oml$, which we would like to measure. We
introduce the factor $c/H_0$ in the definition of $s$ simply in order
to avoid the small figures when we mention 10-100 Mpc scales. This
measure of the radial distance $s$ is a linear extrapolation of the
distance-redshift relation of the nearby universe. We save the unit
$\himpc$ for the usual comoving coordinates, and we use a new notation
$\himpcz$ for the measure $s$. The relation of the distance $s$ and
the comoving distance $x$ is redshift-dependent and is given by a
differential form, $ds = H(z)/H_0\cdot dx $. In the unit system $c=1$
and $H_0 = 1$ which we used in previous sections, the distance $s$ is
simply the redshift $z$.

We put the 1,000 cells in a generic cubic box of $L^3$ on regular
$10\times 10\times 10$ sites in redshift space. The top-hat kernel
[$m=0$ in equation (\ref{eq1-6})] is used and the radius $R$ of the
kernel is just a half of the cell separation: $R = L/20$, so that
there are no overlapping regions to ensure the independence of the
cell volumes. The Fisher matrix is straightforwardly calculated for
this generic cubic region. This generic box is smaller than the SDSS
survey volume, therefore the Fisher matrix can be rescaled as if the
whole survey volume consisted of many independent generic cubic
boxes. One can easily see that the Fisher matrix of the composite
sample is just given by a sum of Fisher matrices of the individual
samples. Although we will miss the information from correlations
between the boxes, the signals from the correlation at such
large separations are very small.

The box size $L$ is chosen so that the cell radius be large enough to
track the linear regime and not to have large shot noise. On the
other hand, too large $L$ makes the cosmic variance large. Therefore,
we should take a smallest $L$ which satisfies the above two
conditions. We take $L=200\himpcz$ for main galaxies and LRG's, and
$L=1000\himpcz$ for quasars. As a result, we have the number
densities $35 /(20\himpcz)^3$ for main galaxies, $0.5 /(20\himpcz)^3$,
for LRG's, and $0.12 /(100\himpcz)^3$ for QSO's.

We assume fiducial values of model parameters such as $(\omm, \oml,
\fbaryon, h, \sigma_8, n) = (0.3, 0.7, 0.13, 0.7, 1, 1)$. The bias
parameters are expected to be different among each samples. We take
fiducial values $b = 1, 2, 2$ for main galaxies, LRG's and quasars,
respectively. For an evaluation of the power spectrum, we use an
useful fitting
formula\footnote{http://background.uchicago.edu/\~{}whu/transfer/transferpage.html}
given by \citet{eis98} instead of running a Boltzmann code such as
the publicly available
CMBfast\footnote{http://physics.nyu.edu/matiasz/CMBFAST/cmbfast.html}
\citep{sel96}.

\begin{figure}[ht]
\epsscale{0.45} \plotone{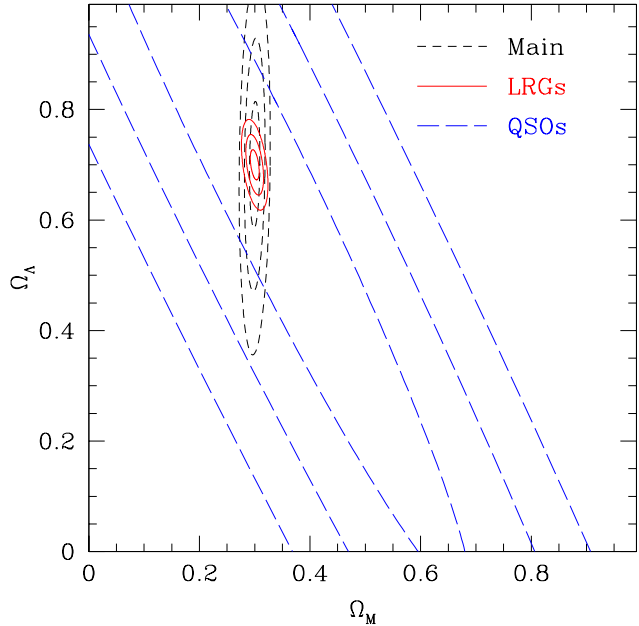} \figcaption[f1.c.eps]{ The
concentration ellipses for the cosmological parameters $\omm$ and
$\oml$. Expected 68\%, 95\%, 99.7\% confidence levels are shown.
Dashed lines are for main galaxies, solid lines for luminous red
galaxies (LRGs), and long-dashed lines for quasars (QSOs).
\label{fig1}}
\end{figure}

In Figure~\ref{fig1} we plot the expected errors from the fiducial
model, when the mass density parameter $\omm$ and the cosmological
constant parameter $\oml$ are simultaneously evaluated.
Three contours for each sample represent the concentration ellipses of
$\nu = 1,2,3$ which correspond to expected $1\sigma$, $2\sigma$,
$3\sigma$ constraints in the parameter space, as described in the
previous section. One can notice that the LRGs impose the best
constraint on the $\Omega_{\rm M}$--$\Omega_\Lambda$ plane. The main
galaxies can only probe a shallow redshift region, so that the
cosmological constant does not strongly contribute to the
redshift-space clustering, and therefore is less constrained. Although
the QSOs can probe deep redshifts, they are too sparse to detect
the redshift-space clustering itself. The LRGs nicely have
intermediate properties and are balanced between the depth and the
density of the survey objects.

\begin{figure}[ht]
\epsscale{0.5} \plotone{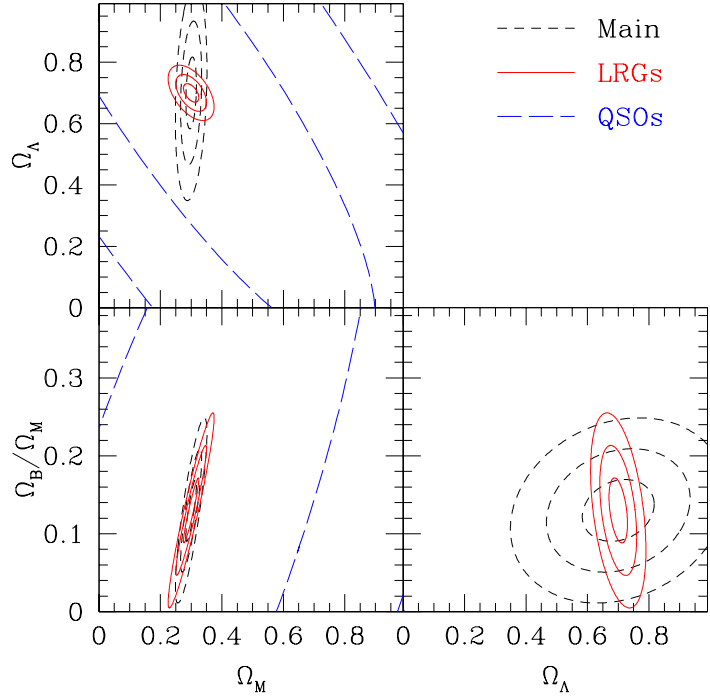} \figcaption[f2.c.eps]{ The
marginalized 2D concentration ellipses when three cosmological
parameters are simultaneously determined.
\label{fig2}}
\end{figure}

In Figure~\ref{fig2} we plot the expected errors when three
parameters, $\omm$, $\oml$, $\fbaryon$ are simultaneously evaluated.
In each panel, marginalized errors are plot for each set of two
parameters, as described in the previous section. In this case, the
QSOs do not constrain each parameter well, because of their
sparseness. Main galaxies and LRGs can impose similar constraints for
the baryon fraction and the mass density parameter.

\begin{figure}[ht]
\epsscale{0.65} \plotone{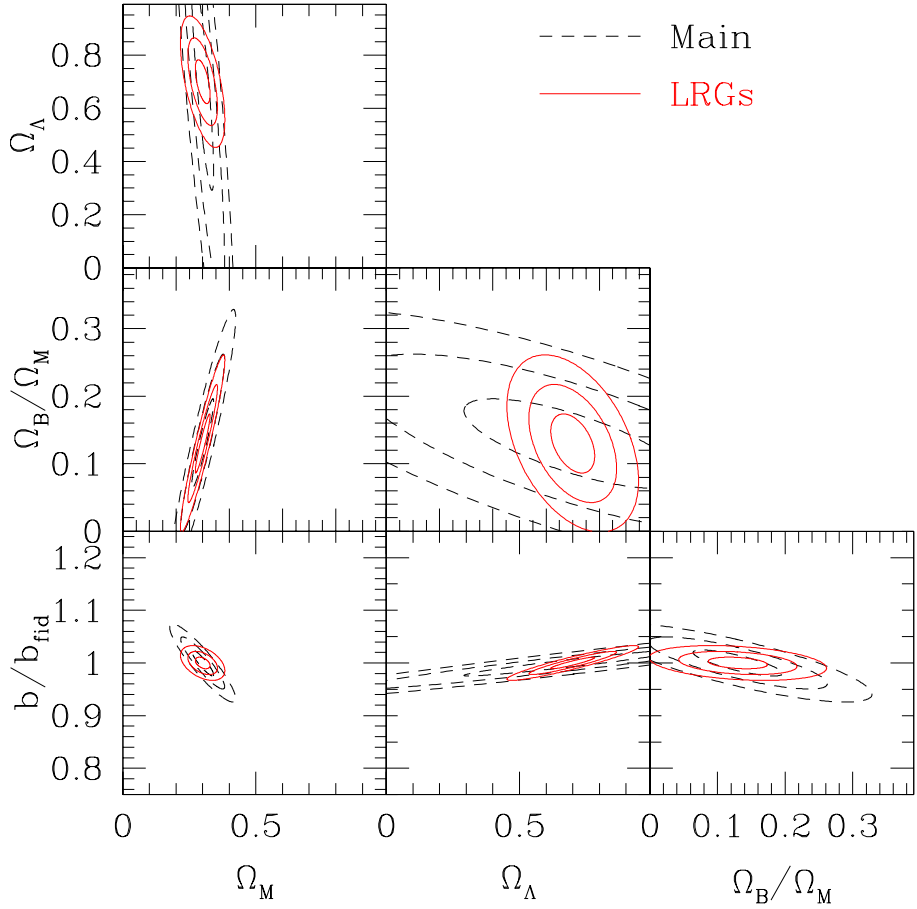} \figcaption[f3.c.eps]{ The
marginalized 2D concentration ellipses when four cosmological
parameters are simultaneously determined.
\label{fig3}}
\end{figure}

In Figure~\ref{fig3} we plot the expected errors when four
parameters, $\omm$, $\oml$, $\fbaryon$, $b$ are simultaneously
evaluated.
The results of the QSOs are not plotted just because the concentration
ellipses are too large and comparable to the bounding boxes. In this
Figure, the bias parameter $b$ is normalized by the fiducial values
$b_{\rm fid} = 1,2,2$ for each sample, just for convenience. The bias
parameters in this Figure are assumed to be constants without
redshift-evolution. If this is not the case, one can divide the sample
into redshift bins, and the expected errors of bias parameter will be
worse, according to the decrease of the sample volume. However, the
error estimates of the other parameters in this Figure are
approximately independent of the redshift evolution of the bias due to
the marginalization over the bias parameter. This is because the
information from cross terms in the redshift bins of $\sim 0.1$ are
already ignored in our analysis, and the likelihood function with a
fixed bias parameter is not too sensitive to the bias parameter
\cite{mat01}.

\begin{figure}[ht]
\epsscale{0.7} \plotone{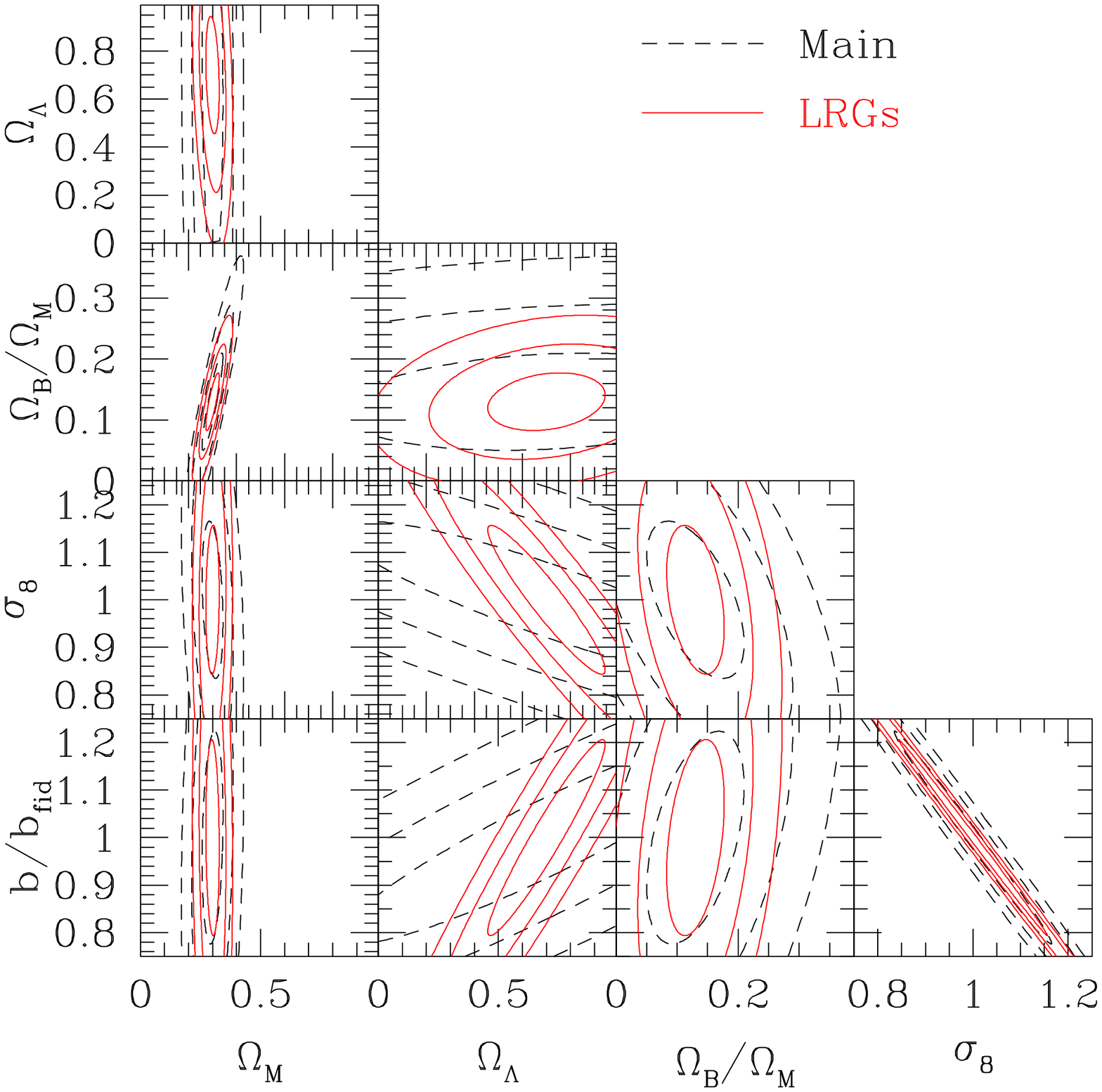} \figcaption[f4.c.eps]{ The
marginalized 2D concentration ellipses when five cosmological
parameters are simultaneously determined.
\label{fig4}}
\end{figure}

\begin{figure}[ht]
\epsscale{0.7} \plotone{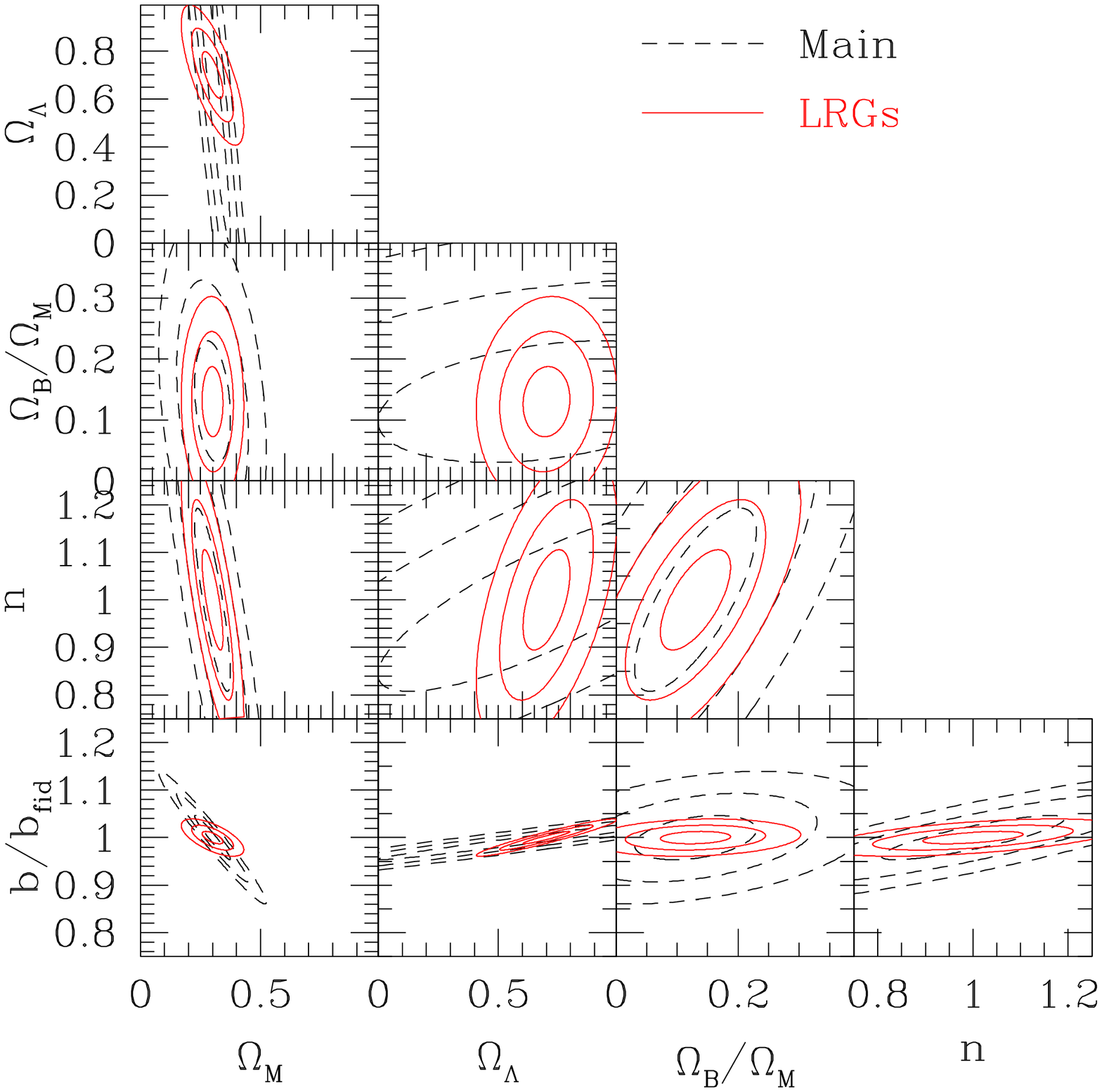} \figcaption[f5.c.eps]{ The
marginalized 2D concentration ellipses when three cosmological
parameters are simultaneously determined.
\label{fig5}}
\end{figure}

In Figure~\ref{fig4} we plot the expected errors, while
simultaneously evaluating five parameters, $\omm$, $\oml$, $\fbaryon$,
$\sigma_8$, and $b$.
The bias parameter $b$ and the normalization $\sigma_8$ are strongly
correlated because their main contribution to the correlation matrix
is through the amplitude of the correlations which is proportional to
$b^2 {\sigma_8}^2$. In Figure~\ref{fig5} we plot the expected
errors, while simultaneously evaluating another set of five parameters,
$\omm$, $\oml$, $\fbaryon$, $n$, and $b$.

Any similar plot can be generated for an arbitrary set of simultaneously
determined parameters. There are
${7\choose1} + {7\choose2} + \cdots + {7\choose7} = 127$ ways to
choose which parameters are fixed, so that we need 122 figures more to
be complete. Instead we have shown only several representative cases
above. In Table~\ref{tab1} are given all the elements of our estimated
seven-dimensional Fisher matrices from which all the information can
be extracted according to the way described in the previous section.
\begin{table}[ht]
\begin{center}
\caption{Normalized seven-dimensional Fisher Information matrices
$F_{\alpha\beta}/(\theta_\alpha \theta_\beta)_{\rm fid}$ 
for the SDSS main galaxies, LRG's, and QSO's. \label{tab1}}
\begin{tabular}{c|ccccccc}
\tableline\tableline
Main       & $\omm$ & $\oml$ & $h$ & $\fbaryon$ & $n$ & $\sigma_8$ &
        $b$ \\
\tableline
$\omm$     &  $1040.$ & $-25.8$ & $675.$ & $-161.$ &  $1050.$ &
 $4080.$ & $ 3200.$ \\ 
$\oml$     & $-25.8$ &  $37.9$ & $2.38$ & $-0.0284$ & $14.0$ &
$-1090.$ & $-989.$ \\ 
$h$        &  $675.$ & $2.38$ &  $619.$ & $-148.$ &  $995.$ &
 $978.$ &  $818.$ \\ 
$\fbaryon$ & $-161.$ &  $-0.0284$ & $-148.$ &  $36.0$ & $-235.$ &
$-262.$ & $-219.$ \\ 
$n$        &  $1050.$ & $14.0$ &  $995.$ & $-235.$ &  $1640.$ &
$938.$ &  $792.$ \\ 
$\sigma_8$ &  $4080.$ & $-1090.$ &  $978.$ & $-262.$ &  $938.$ &
$51600.$ & $43200.$ \\ 
$b$        &  $3200.$ & $-989.$ &  $818.$ & $-219.$ &  $792.$ &
$43200.$ & $36600.$ \\
\tableline\tableline
LRG's      & $\omm$ & $\oml$ & $h$ & $\fbaryon$ & $n$ & $\sigma_8$ &
        $b$ \\
\tableline
$\omm$     &  $1740.$ &  $513.$ &  $1680.$ & $-398.$ &  $2710.$ &
$-2400.$ & $-2310.$ \\
$\oml$     &  $513.$ &  $804.$ &  $360.$ & $-81.8$ &  $653.$ &
$-7420.$ & $-6850.$ \\
$h$        &  $1680.$ &  $360.$ &  $1730.$ & $-418.$ &  $2760.$ &
 $-1510.$ &  $-1370.$ \\
$\fbaryon$ & $-398.$ & $-81.8$ & $-418.$ &  $103.$ & $-661.$ &
$308.$ & $280.$ \\
$n$        &  $2710.$ &  $653.$ &  $2760.$ & $-661.$ &  $4490.$ &
 $-3390.$ &  $-3060.$ \\
$\sigma_8$ & $-2400.$ & $-7420.$ &  $-1510.$ & $308.$ &  $-3390.$ &
$88400.$ & $78700.$ \\
$b$        & $-2310.$ & $-6850.$ &  $-1370.$ & $280.$ &  $-3060.$ &
$78700.$ &  $70500.$ \\
\tableline\tableline
QSO's      & $\omm$ & $\oml$ & $h$ & $\fbaryon$ & $n$ & $\sigma_8$ &
        $b$ \\
\tableline
$\omm$ & $10.2$ & $11.5$ & $9.28$ & $-2.38$ & $14.5$ & $-27.5$ &
$-23.5$ \\
$\oml$ & $11.5$ & $13.7$ & $10.4$ & $-2.60$ & $16.3$ & $-31.5$ &
$-27.5$ \\
$h$ & $9.28$ & $10.4$ & $8.72$ & $-2.28$ & $13.5$ & $-24.2$ &
$-20.8$ \\
$\fbaryon$ & $-2.38$ & $-2.60$ & $-2.28$ & $0.614$ & $-3.50$ &
$6.00$ & $5.16$ \\ 
$n$ & $14.5$ & $16.3$ & $13.5$ & $-3.50$ & $21.0$ & $-38.3$ &
$-32.9$ \\
$\sigma_8$ & $-27.5$ & $-31.5$ & $-24.2$ & $6.00$ & $-38.3$ &
$78.1$ & $66.7$ \\
$b$ & $-23.5$ & $-27.5$ & $-20.8$ & $5.16$ & $-32.9$ & $66.7$ &
$57.4$ \\
\tableline
\end{tabular}
\end{center}
\end{table}
The listed matrix elements are normalized by fiducial values as
$F_{\alpha\beta}/(\theta_\alpha \theta_\beta)_{\rm fid}$. Thus, the
inverse of square-root of each diagonal element gives the fractional
error relative to the fiducial value of each parameter when a
single parameter is determined knowing the other parameters. The
square-root of the diagonal elements of the inverse of any sub-matrix
gives the fractional error of the parameters when corresponding
parameters are simultaneously determined.

Basically, the LRG's can impose constraints on each parameter
comparable to, or even better than main galaxies. The reason that the
main galaxies do not work so well (even though their number is 10
times larger than that of LRG's) is that we only use the linear regime
of the clustering. Therefore, the depth of the LRG sample is more
critical for the parameter estimation than the higher density of the
main galaxies. The quasars are hopeless when many parameters are
simultaneously determined, because of their extreme sparseness. In
this way, our analysis gives the idea how the dense sampling affects
cosmological parameter estimations.

\section{Conclusions}

In this paper, we described a maximum-likelihood method to determine
the cosmological parameters from apparent redshift-space clustering.
We took advantage of the analytical formula for linear two-point
correlations in redshift space with cosmological distortions.  We
introduced an accurate approximation in calculating the elements of
the correlation matrix among cells in redshift-space. This step makes
the method practically feasible in spite of the large dimensionality of
the correlation matrices.

In the maximum-likelihood method, one can easily evaluate the expected
parameter estimation errors in any sample from the Fisher information
matrix. We have calculated the seven-dimensional Fisher matrix for
three types of objects in the SDSS survey: main galaxies, LRGs and
quasars. To illustrate the behavior of the multi-dimensional Fisher
matrix we have used concentration ellipses in marginalized
two-dimensional parameter space.

In this paper, we divided the survey volume into generic boxes in
order to simplify the Fisher matrix estimation. We ignored the
correlations between these sub-regions, so the constraints will
improve somewhat if those correlations are properly included. However,
the inversion of the resulting huge matrices can become extremely
time-consuming. The use of the Karhunen-Lo\`eve (KL) transform is a
practical strategy in this case. Such methods can also be used in a
targeted data-compression role to find linear combinations of counts
which retain as much information about the parameters as possible
\citep{vog96,teg97,mat00b,tay01}.

The choice of the cell radius is somewhat arbitrary in this work. We
choose the spherical cell with radius of $10\himpc$ for galaxies and
LRGs, which is the border of the linear regime. With larger cell
radius, the validity of the linear theory increases and the shot noise
is reduced. The cosmic variance, however, increases with cell radius.
The parameter estimation is dominated by the highest signal-to-noise
modes, which are at large wavelength, in particular for the case of
the LRG sample. The high frequency modes close to pixel scales mostly
contain shot noise after the KL transformation. As a result, we
believe our conclusions are not sensitive to the choice of the cell
radius. A fully accurate determination of the optimal choice of the
cell radius depends on the behavior of the nonlinear effects, so that
a comparison with numerical simulation is needed, beyond the scope of
the current work.

We have considered three subsets of the SDSS redshift data, spanning a
wide range of depth, sampling density and intrinsic clustering
strength.  We found, that for measuring cosmological parameters in the
linear regime there is a clear optimum, represented by the
intermediate-redshift LRG's.  The low spatial density of quasars is
not overcome by their much larger depth, and the relatively small
depth of the main SDSS galaxies is not compensated by their high
sampling density -- the redshift is not high enough to test curvature,
and their cosmic variance is too large. The LRG sample, much smaller in
numbers than the main sample, and much shallower than the quasars is
an excellent compromise between sampling density and cosmological
depth. The constraints derived from the LRGs are much tighter than for
the other two samples.

The advantage of these intermediate-redshift objects, and the logic
behind this optimum goes beyond the SDSS. In designing future redshift
surveys, it is important to find the right balance between the density
of objects and the survey depth. Their interplay can be quite complex,
as we have shown here. The relation between accuracy and sky coverage
is simple and can be estimated analytically.


\acknowledgements

We would like to acknowledge useful discussions with Dan VandenBerk,
Daniel Eisenstein and Adrian Pope. TM acknowledges support from the
Ministry of Education, Culture, Sports, Science, and Technology,
Grant-in-Aid for Encouragement of Young Scientists, 13740150, 2001. AS
acknowledges support from grants NSF AST-9802 980 and NASA LTSA
NAG-53503.

\clearpage


\clearpage

\end{document}